\documentclass[twocolumn,prl,aps,amsmath,superscriptaddress,showpacs]{revtex4}

\usepackage{epsfig}
\usepackage{float}

\def\gee        {\epsilon}

\def\go         {\omega}

\def\gt         {\theta}

\def\ra         {\rangle}
\def\nk         {n{\bf k}}

\def\npkmq      {n'\({\bf k-q}\)}
\def\dk         {\frac{d\,{\bf k}}{\(2 \pi\)^3}}
\def\dk1        {\frac{d\,{\bf k}_1}{\(2 \pi\)^3}}

\renewcommand{\[}{\left[}
\renewcommand{\]}{\right]}
\renewcommand{\(}{\left(}
\renewcommand{\)}{\right)}

\restylefloat{figure}

\begin{document}
\title{Quasiparticle bandstructure effects on the {\it d}
hole lifetimes of copper within the $GW$ approximation}
\author{Andrea Marini}
\affiliation{
Istituto Nazionale per la Fisica della Materia,
Dipartimento di Fisica dell'Universit\`a di Roma ``Tor Vergata'', Italy.}
\affiliation{
Departamento de F\'\i sica de Materiales,
Facultad de Ciencias Qu\'\i micas,
Universidad del Pais Vasco, Centro Mixto USPV/EHU--CSIC and  Donostia International Physics Center.
E--20018 San Sebasti\'an, Basque Country, Spain}
\author{Rodolfo Del Sole}
\affiliation{
Istituto Nazionale per la Fisica della Materia,
Dipartimento di Fisica dell'Universit\`a di Roma ``Tor Vergata'', Italy.}
\author{Angel Rubio}
\affiliation{
Departamento de F\'\i sica de Materiales,
Facultad de Ciencias Qu\'\i micas,
Universidad del Pais Vasco, Centro Mixto USPV/EHU--CSIC and  Donostia International Physics Center.
E--20018 San Sebasti\'an, Basque Country, Spain}
\author{Giovanni Onida}
\affiliation{
Istituto Nazionale per la Fisica della Materia,
Dipartimento di Fisica dell'Universit\`a di Roma ``Tor Vergata'', Italy.}
\affiliation{
Istituto Nazionale per la Fisica della Materia, 
Dipartimento di Fisica dell'Universit\'a di Milano, Italy.}
\date{\today}

\begin{abstract}
 We investigate the
 lifetime of {\it d} holes in copper within a {\it first--principle} $GW$ approximation.
 At the $G_0W_0$ level the lifetime of the topmost {\it d} bands are
 in agreement with the experimental results and are
 four times smaller than those obtained in the ``on--shell'' calculations
 commonly used in literature.  The theoretical
 lifetimes and bandstructure, however, worsen when further iterative 
 steps of self-consistency are included in the calculation, pointing to a
 delicate interplay between self--consistency and the inclusion of 
 vertex corrections.
 We show that the $G_0W_0$ success in the lifetimes calculation is due to the opening of new
 ``intraband'' decay channels that disappear at self--consistency.
\end{abstract}

\maketitle

\noindent Very recent experimental and theoretical results on
the quasiparticle lifetimes in
noble~\cite{campillo}--\cite{ferdi}  and simple~\cite{silkin} metals 
show that our present understanding of the electron dynamics in real solids
is far from being complete~\cite{angelrev}.
For electrons above the Fermi level (hot electrons)
time--resolved two--photon photoemission experiments show a non quadratic behavior of
the lifetime, in contrast with the Fermi liquid theory prediction~\cite{graphite}.
For occupied states, the number of possible scattering
events increases rapidly as their energies decrease below the Fermi level,
and the agreement between photoemission experiments and theoretical results worsens~\cite{angel}.
An open question is whether this discrepancy comes from 
effects beyond the approximation used in the calculations 
(i.e., beyond $GW$, where vertex corrections are neglected),
or it comes from the way used to solve
the quasiparticle equation for a given self--energy. In the present work
we address quantitatively the latter point, the inclusion of
vertex corrections  being beyond the scope of this paper.

Density Functional Theory~\cite{DFT} (DFT) has become the
state--of--the--art approach to study ground state properties of a large
number of systems, going from molecules, to surfaces, to complex solids
\cite{dft-app}.
The success of DFT is based on the idea that the spatial density of a system of 
interacting particles can be exactly described by a non--interacting gas 
of  Kohn-Sham (KS) independent particles, moving under the action of an effective potential 
which includes the exchange-correlation potential $V_{xc}$. 
Compared with experiment, the usual local density (LDA)~\cite{LDA} approximation to the
DFT yields semiconductor bandstructure  which systematically
underestimate the band gap, while for noble metals the discrepancies are both 
{\bf k}--point and band dependent~\cite{noi}.
Another important drawback of the use of DFT eigenvalues as excitations (bandstructure)
energies is that they are by construction real; no lifetime
effects are included. 
An alternative approach is Time Dependent DFT where
all neutral excitations are, in principle, exactly described~\cite{rmp}.
However, quasiparticle lifetimes have not been considered so far within this approach.

Many--body perturbation theory allows one to
obtain band energies and lifetimes in a rigorous way, i.e. as the poles of
the one--particle Green's function $G$~\cite{rmp}. 
Those are determined by the solution of a Dyson-like equation of the form
\cite{hedin}:
\begin{multline}
 \[-\frac{\hbar^2}{2m}\nabla^2_{{\bf r}}+ V_{external}\({\bf r}\)+
  V_{Hartree}\({\bf r}\)\]\psi_{\nk}\({\bf r},\go\)+\\
 \int\,d{\bf r}'\,\Sigma\({\bf r}\,{\bf r}',\go\)\psi_{\nk}\({\bf r}',\go\)=
 E_{\nk}\(\go\)\psi_{\nk}\({\bf r},\go\),
 \label{eq1}
\end{multline}
containing the  non-local, generally complex and non-hermitian, frequency dependent
self--energy operator $\Sigma$.
The poles of G are the quasiparticle (QP) energies $\gee_{\nk}^{QP}$, that from Eq.\,(\ref{eq1}) 
correspond to the generally complex solutions of the equation:
$\gee_{\nk}^{QP}=E_{\nk}\(\gee_{\nk}^{QP}\)$.
The real part of $\gee_{\nk}^{QP}$ gives the quasiparticle bandstructure; whereas the imaginary part yields the inverse quasiparticle lifetime.
In the present work $\Sigma$ is 
evaluated according to the so-called GW approximation,
derived by Hedin in 1965~\cite{hedin}, which is
based on an expansion in
terms of the dynamically screened Coulomb interaction $W\(\go\)$. In the first iteration
$G_0$ and $W_0$ from DFT calculations are used to compute $\Sigma$ as $\Sigma_0=iG_0W_0$.
Unlike semiconductors, the case of noble metals has been studied only 
recently~\cite{campillo,noi,campillo2}.
The lifetime of hot--electrons in copper has been calculated with the $GW$ self--energy
evaluated at the DFT zero--order energies~\cite{campillo} 
(namely ``on mass-shell'' approximation). This
approach is based on the assumption of vanishing QP corrections of the DFT states
while, very recently, large QP effects on the occupied bands of copper have been
found~\cite{noi}.
The ``on mass-shell'' approach applied to the hole lifetimes~\cite{campillo2}
indicates that d--holes in copper exhibit a longer lifetime than excited s/p electrons.
The quantitative comparison with experiment~\cite{angel}, however, has shown
a large overestimation of the experimental lifetimes measured by means of photoemission 
spectroscopy.
In this paper we go beyond the ``on mass-shell'' approximation, and 
calculate the lifetimes of {\it d} bands in copper fully solving the 
QP Eq.\,(\ref{eq1}) in the complex plane without relying on
any analytic continuation. The convergence of the results is carefully
checked.  Our results are significantly different from those
obtained within a DFT self--energy based approach and are in good agreement with
experiments. The careful analysis of the physics underlying the GW
decay of quasiholes will
shed light into the quantum--mechanical mechanism behind the electron dynamics in noble metals.

\begin{figure}[H]
\begin{center}
\epsfig{figure=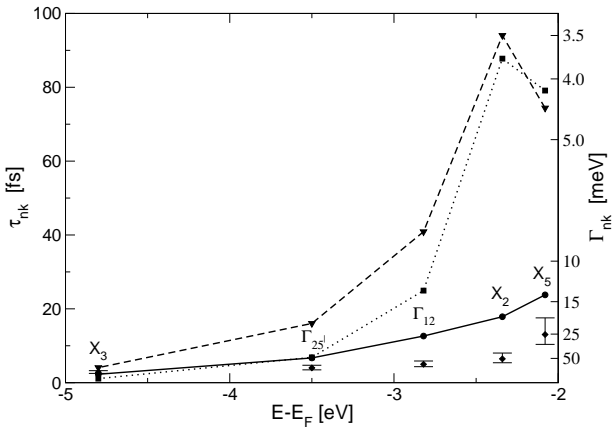,clip=,bbllx=55,bblly=0,bburx=594,bbury=800,angle=-90,width=8.5cm}
\end{center}
\vspace{-.5cm}
\caption{
\footnotesize{
Lifetimes of selected {\it d}-bands of copper.
Diamonds with error bars: experimental data from Ref.~\cite{angel}. The theoretical
results obtained in this work are reported for different level of iterations within the
the $G_iW_0$ quasiparticle
approximation. Full line; $G_0W_0$. Dotted line: ``on mass-shell'' $G_0W_0$.
Dashed line:  ``on mass-shell'' $G_1W_0$. All theoretical quasiparticle energy position are aligned
to the corresponding experimental values. 
For a comparison of the band positions see Tab.\,\ref{tab2}.
The lifetimes are systematically above the experimental values, as expected for the
contribution of higher--order electron--electron and residual phononic contributions.
}}
\label{fig1}
\end{figure}

In our approach, we start by solving Eq.\,(\ref{eq1}) on the
real axis, as it is commonly done in QP bandstructure calculations~\cite{gwrev}.
In this way we obtain $\gee_{\nk}^{QP,0}$, a first guess for the real QP energies:
\begin{align}
 \gee_{\nk}^{QP,0}=\gee_{\nk}^{DFT}+Re\[\Sigma_{\nk}\(\gee_{\nk}^{QP,0}\)\]-V^{\nk}_{xc},
 \label{eq2} 
\end{align}
where $\Sigma_{\nk}\(\go\)\equiv\langle\nk|\Sigma\({\bf r}\,{\bf r}',\go\)|\nk\rangle$ and
$V^{\nk}_{xc}\equiv\langle\nk|V_{xc}\({\bf r}\)|\nk\rangle$.
Even if Eq.\,(\ref{eq1}) in principle requires a diagonalization with respect to the band index $n$,
it can be reduced to the form of Eq.\,(\ref{eq2}) because
the computed off--diagonal matrix elements of $\Sigma$ are at least two orders of magnitude smaller
than the diagonal ones 
($\left|\Sigma_{n,n'}\right|<<\left|\Sigma_{n,n}\right|$ $\forall\,n,n'$).
The difference between the requested exact quasiparticle energy
$\gee_{\nk}^{QP}$  and the first guess
defined in Eq.\,(\ref{eq2}) is due to the fact that the self--energy
has an imaginary part, namely:
\begin{align}
 \gee_{\nk}^{QP}-\gee_{\nk}^{QP,0}=i\,Z_{\nk} Im\[\Sigma_{\nk}\(\gee_{\nk}^{QP,0}\)\],
 \label{eq3} 
\end{align}
and
\begin{align}
 Z_{\nk}\equiv\[1-\left.\frac{d\,\Sigma_{\nk}\(\go\)}{d\,\go}\right|_{\go=\gee_{\nk}^{QP,0}}\]^{-1}.
 \label{eq4}
\end{align}
Since the quasiparticle concept holds when $Im\[\Sigma_{\nk}\(\gee_{\nk}^{QP,0}\)\]$ 
is small $\gee_{\nk}^{QP,0}$  is a natural starting
point to get the QP excitation.
Note that being $Z_{\nk}$ complex Eq.\,(\ref{eq3}) will also slightly modify the real 
part of $\gee_{\nk}^{QP,0}$.
Thus the corresponding linewidth ($\Gamma_{\nk}$) and lifetime ($\tau_{\nk}$) are given by
\begin{align}
\tau_{\nk}^{-1}\equiv 2\Gamma_{\nk}\equiv 2 Re\[Z_{\nk}\]Im\[\Sigma_{\nk}\(\gee_{\nk}^{QP,0}\)\].
\label{eq4a}
\end{align}
A remarkable property of Eq.\,(\ref{eq4a}) is that,
being $\gee_{\nk}^{QP,0}$ solution of Eq.\,(\ref{eq3}), only the $Re\[Z_{\nk}\]$ is needed 
to define the QP lifetime in contrast with the common expansion
of $\Sigma_{\nk}\(\go\)$ around the DFT energy $\gee_{\nk}^{DFT}$.
$Z_{\nk}$ is the usual renormalization factor, referred to the initial
QP guess instead of to the DFT eigenvalue.
A first approximate solution of Eqs.\,(\ref{eq1}--\ref{eq2})
can be obtained by fully neglecting the QP
correction $Re\[\Sigma_{\nk}\(\gee_{\nk}^{QP,0}\)\]$, i.e., by 
assuming  $\gee_{\nk}^{QP,0}\equiv\gee_{\nk}^{DFT}$
and hence $Z_{\nk}=1$.  
The corresponding lifetime is 
$\tau_{\nk}\approx \left\{2 Im\[\Sigma_{\nk}\(\gee_{\nk}^{DFT}\)\]\right\}^{-1}$
and is usually referred in the literature as the
``on mass-shell'' $G_0W_0$ lifetime~\cite{quinn,angelrev},
because the input energies  of the QP
equation are supposed to remain constant, and are subsequently used to calculate the lifetimes.
Another approach
uses Eq.\,(\ref{eq3}) with {\it real} $Z_{\nk}$, and $\gee_{\nk}^{QP,0}$
corresponding to a LMTO--$GW$ bandstructure~\cite{ferdi}.
This method has confirmed the overestimation of the top d--bands lifetimes found 
in the ``on mass-shell'' approximation~\cite{campillo2}. However, we stress
that the approximated real QP energies $\gee_{\nk}^{QP,0}$ used in Ref.~\cite{ferdi}
{\it are not} solution of Eq.\,(\ref{eq2}).  As a consequence, $\Gamma_{\nk}$, obtained from 
Eq.\,(\ref{eq3}), contains and additional term proportional to 
$Im\[Z_{\nk}\]$ that usually is neglected~\cite{ferdi} or, it is very small.
However for the case of noble metals it can be as large as $20$\,meV for
the top d--bands (see  Fig.~\ref{fig1}), accounting for 10--40\% of the total electronic linewidth.
This term  reduces the lifetimes, in agreement with the experiment and with the results presented
below.

Al the present calculations of Green's function and screened interaction have been 
performed using a plane wave basis.
We have used norm--conserving pseudopotentials~\cite{MT,CUPRB} in the DFT--LDA calculation, 
with a 60\,Ry energy cutoff, corresponding to $\sim\,800$ plane waves~\cite{notaconv}.
A particular care has been devoted to the check of the effect of the Lorentzian
broadening $\eta$ included in the screening function $W\(\go\)$ for numerical reasons
(see Ref.~\cite{gwrev} for details). Calculations
with different broadening values  have shown an increase up to $50\%$ of the lifetimes
when $\eta$ is reduced from $0.1$ to $0.005$\,eV. 
This dramatic numerical effect can be avoided by using $\eta=0.005$\,eV, yielding results which
coincides with those extrapolated at $\eta=0$\,eV. This is an important point
since most calculations presented so far have been done using dampings of
0.1 eV or larger.

In Fig.~\ref{fig1} we present the ``on mass-shell'' $G_0W_0$ result compared with 
the full solution of Eqs.\,(\ref{eq2}--\ref{eq3})  (simply referred as to  $G_0W_0$), 
and with the experimental photoemission results~\cite{angel}.
The ``on mass-shell'' $G_0W_0$ yields lifetimes which are
four times too large at the {\it d}--bands top, and
three times too small at the {\it d}--bands bottom (not shown in Fig.~\ref{fig1}).
The $G_0W_0$ results, instead, are in rather good
agreement with experiments. 
Actually, they are systematically above the experimental values, as expected for the
contribution of higher--order electron--electron and residual phononic and
impurity contributions.
Moreover, as shown in Tab.~\ref{tab2}, the energy positions of the quasiparticle peaks are well reproduced
in $G_0W_0$, while in the ``on mass-shell'' calculation  the eigenvalues
(coincident with DFT ones) span a larger range 
of energies, reflecting the known DFT overestimation of the linewidth of {\it d}--bands~\cite{noi}.
The origin of the big difference between the two lifetime 
calculations stems from the large QP self--energy corrections
to the {\it d}--bands of copper, completely neglected in ``on mass-shell'' calculations.
\begin{table}[H]
\begin{center}
\begin{tabular}{cccccc} \hline \hline
& DFT  &  $G_0W_0$ & $G_1W_0$   & $G_2W_0$ &  Experiment  \\ \hline
$\Gamma_{12}-\Gamma_{25'}$    & $0.91$  & $0.60$  & $0.38$  & $0.23$  & $0.81$ \\
$X_5-X_3$     & $3.23$  & $2.49$  & $1.99$  & $1.65$  & $2.79$ \\
$X_5-X_1$     & $3.70$  & $2.90$  & $2.31$  & $1.92$  & $3.17$ \\
$L_3-L_3$     & $1.58$  & $1.26$  & $1.03$  & $0.90$  & $1.37$ \\
$L_3-L_1$     & $3.72$  & $2.83$  & $2.13$  & $1.65$  & $2.91$ \\ \hline
$L_1-L_{2'}$  &  $5.40$ & $4.76$  & $4.78$  & $3.77$  & $4.95$ \\ \hline \hline
\end{tabular}
\end{center}
\caption{
\footnotesize{Theoretical bandwidths (in eV) and band energies
for copper, at high-symmetry points and for various iterations of the $G_iW_0$ quasiparticle
approximation.
There is a striking agreement with the experiment at the $G_0W_0$ level~\cite{noi},
but this worsens when the number of iterations is increased, 
showing the  potential importance of including also vertex corrections.
The experimental values are taken from ref.~\cite{courths}.}}
\label{tab2}
\end{table}

The proper inclusion of these non trivial self--energy corrections change the electronic decay channels of
the {\it d} levels. In particular, we expect a strong effect on the topmost bands.
Let us develop this idea further, by looking at the $G_0W_0$ expression for the self--energy:
\begin{align}
\Sigma_0\({\bf r}\,{\bf r}',\go\) = 
 \int\,\frac{i\,d\go'}{2\pi}G_0\({\bf r}\,{\bf r}',\go'\)W_0\({\bf r}\,{\bf r}',\go-\go'\).
\label{eq5a}
\end{align}
$W_0\({\bf r},{\bf r}';\go'\)$  is the dynamically screened potential (convolution
of the inverse dielectric function $\gee^{-1}\({\bf r},{\bf r}';\go'\)$
with the bare Coulomb potential).
We can write explicitly Eq.\,(\ref{eq3}) in terms of a summation over the DFT states
embodied in $G_0$ as:
\begin{multline}
\Gamma_{\nk} \propto 
 \sum_{n'}\sum_{\bf q} Im\[W_{\nk\rightarrow\npkmq}\(\gee_{\npkmq}^{DFT}-\gee_{\nk}^{QP,0}\)\] \\
 \gt\(\gee_{\npkmq}^{DFT}-\gee_{\nk}^{QP,0}\)f_{\npkmq},
\label{eq6} 
\end{multline}
where
\begin{multline}
W_{\nk\rightarrow\npkmq}\(\go\)\equiv
\int\,d{\bf r}d{\bf r}' \psi^*_{\nk}\({\bf r}\) \psi_{\npkmq}\({\bf r}\)\\
 W\({\bf r}\,{\bf r}',\go\) \psi^*_{\npkmq}\({\bf r}'\) \psi_{\nk}\({\bf r}'\),
\label{eq6a} 
\end{multline}
being $f_{\nk}$ the Fermi occupations and $\gt\(\go\)$ the step function.
In Eq.\,(\ref{eq6}) a quasihole in the state $|\nk\ra$ looses energy  via transitions
to all the possible occupied states with (higher) energy $\gee_{\npkmq}^{DFT}$; 
the energy difference is dissipated by the
the screening cloud (described by $W\(\go\)$) that surrounds the DFT hole $|\nk\ra$.
Thus the $G_0W_0$ transitions contributing to the hole linewidth look like single particle
transitions from a quasihole to a DFT hole.
In higher order of Hedin's equations, i.e. including vertex corrections, this
interpretation of transitions looses meaning.
This is due to
self--energy effects on the Green's function of Eq.\,(\ref{eq5a}) and
interactions of the screening cloud with the state $|\npkmq\ra$ (vertex corrections~\cite{hedin})
that in Eq.\,(\ref{eq6}) are neglected. 

For the  top of the occupied {\it d} bands the $G_0W_0$ calculation yields negative 
QP corrections~\cite{noi}; this means that Eq.\,(\ref{eq6})
contains also contributions coming from the decay of the QP  {\it d} band to {\it exactly the same} 
DFT band. We will refer to these transitions as ``intraband'' decay channels. 
These contributions are important because the {\it d}--bands of copper are flat, hence the corresponding
density of states is large.
Moreover the screened interaction between {\it d}--bands is strong, as the  {\it d}--states are 
spatially localized and screening is less effective at small distances.
In the ``on mass-shell'' calculation the states $|\npkmq\ra$ appearing in Eq.\,(\ref{eq6}) and
the quasiparticle states correspond to the same DFT eigenvalues; 
this means that the ``intraband'' decay channels occur at zero energy,
where the low--energy Drude tail of the dielectric function dominates (and hence
the screened interaction is vanishing).
This leads to the usual interpretation of the long calculated lifetimes at the {\it d} bands top
as due to the fact that these {\it d} states can only  decay to {\it s/p} states.
These matrix elements are smaller than those involving states with the same {\it l}--character.

Even if the results of the $G_0W_0$ calculation are in rather good agreement with the experiments, 
a natural question about
the physical meaning of the ``intraband'' decay  channels arises. Being not at self-consistency,
the system is described within $G_0W_0$ on the basis of quasiparticle states and DFT states 
(those involved
in the hole decay)  with different energies. To remove this inconsistency Eq.\,(\ref{eq1}) should be 
solved iteratively, until converged QP energies are obtained.
However, as we show below, iterations beyond the usual $G_0W_0$  level worsen both
the imaginary and real  parts of the calculated QP energies. 

So far fully self-consistent $GW$ calculations have been performed only for the homogeneous
electron gas~\cite{holm} and for simple semiconductors and metals~\cite{eguiluz},
yielding worse spectral properties than those obtained in the non self-consistent $G_0W_0$.
The construction of a self-consistent $GW$ self--energy is a formidable task even for the simple
systems mentioned above.
In copper, already  the update of the screening function is 
rather demanding, due to the presence of  localized {\it d}  orbitals that imply a large cut--off in the
plane wave expansion.
To test the effect of self-consistency on the QP energies  and lifetimes we use a simplified $G_iW_0$ method,
where the self--energy operator is defined as
\begin{align}
\Sigma_{i}\({\bf r}\,{\bf r}',\go\)\equiv
 \int\frac{i\,d\go'}{2\pi}G_{i}\({\bf r}\,{\bf r}',\go'\) W_{\(0\)}\({\bf r}\,{\bf r}',\go-\go'\),
\label{eq7} 
\end{align}
being  $i$ is the iteration number. $G_{i}$ involves 
the QP energies obtained  from $\Sigma_{\(i-1\)}$, without considering renormalization
factors, lifetimes, and energy structures beyond QP peaks.
As the QP bandstructure resulting  from the first iteration is already in excellent agreement
with experiment, our next step is to perform an ``on mass-shell'' $G_1W_0$ calculation.
The resulting lifetimes are compared with experiment in Fig.\ref{fig1}. One sees that forcing
QP energies to appear also in the states involved in Eq.\,(\ref{eq6}),
i.e. the states forward which the quasihole decays, yields
lifetime results similar to those of the ``on mass-shell'' $G_0W_0$
method. The same overestimation of the lifetimes of the top of the {\it d} bands is observed, 
confirming that a good agreement with experiment depends on the inclusion  of the 
``intraband''  decay channels described by Eq.\,(\ref{eq6}).

A further question could now be addressed and is
 how self-consistency affects the 
quasiparticle bandstructure obtained within $G_0W_0$.
As shown in Tab.~\ref{tab2}, at higher iteration
orders of the quasiparticle $GW$ equation 
the resulting energies worsen. 
The {\it d}--bands width decreases, reducing the agreement with experiment. 

These results show that $G_0W_0$ describes correctly the {\it d}--holes lifetimes as far as
``intraband''  decay channels between {\it d}--like states are included.
Those transitions, however, imply an inconsistency between the quasiparticle initial states and the
DFT final states of the hole decay, as shown in Eq.\,(\ref{eq6}).
A self--consistent solution of Dyson equation removes this inconsistency, worsening, however,
the agreement with the experimental results.
This is a clear indication of the need of including vertex corrections in the self--consistent
procedure. As shown for simple metals~\cite{shirley}, vertex corrections would partially
cancel the dressing of the $G_0$ Green's function of Eq.\,(\ref{eq5a}), restoring the
``intraband''  decay channels and, consequently, the $G_0W_0$ results.

In conclusion, we have shown that the lifetimes of {\it d}--holes in copper, calculated
within the $G_0W_0$ method, are in good agreement with the experimental results, and can
be obtained within the very same scheme which yields a good quasiparticle bandstructure.
In contrast, further iterations of the QP equation beyond the $G_0W_0$  level yield worse results,
for both the real and imaginary parts of self--energy.
This can be explained by the need of including also vertex corrections together with 
self--consistency. This result is quite general and should apply to all metals in which one has two
or more sets of electronic states with different degrees of spatial localization.

This work has been supported by the INFM  PRA project ``1MESS'', MURST-COFIN 99,
Basque Country University, Iberdrola S.A. and DGES. and
by the EU through the NANOPHASE Research Training Network
(Contract No. HPRN-CT-2000-00167). We thank F. Aryasetiawan and P.M. Echenique for
helpfull discussions.


\end{document}